\documentclass[useAMS,usenatbib,usegraphicx,footnotetext]{mn2e}

\usepackage{multirow}

\newcommand{\hii}{H\textsc{ii}}

\def\ks{km s$^{-1}$}

\def\m{$^\prime$}
\def\s{$^{\prime\prime}$}

\def\cm3{cm$^{-3}$}

\def\2{$^{12}$CO}
\def\3{$^{13}$CO}
\def\H{HCO$^{+}$}
\def\msun{M$_\odot$}

\def\pap{Paper\,I}

\title[The ISM towards two pillar-like features]{The molecular environment of the pillar-like features 
in the HII region G46.5-0.2}

\author[S. Paron et al.]{ {\Large S. Paron$^{1,2}$\thanks{sparon@iafe.uba.ar},  
M. Celis Pe\~{n}a$^{1}$, 
M. E. Ortega$^{1}$, 
C. Fari\~{n}a$^{3,4}$,  
A. Petriella$^{1}$, 
M. Rubio$^{5}$ 
and R. P. Ashley$^{3,6}$}
\\ 
$^{1}$ CONICET-Universidad de Buenos Aires. Instituto de Astronom\'{\i}a y F\'{\i}sica del Espacio
             CC 67, Suc. 28, 1428 Buenos Aires, Argentina \\
$^{2}$ Universidad de Buenos Aires. Facultad de Arquitectura Dise\~{n}o y Urbanismo. Buenos Aires, Argentina\\
$^{3}$ Isaac Newton Group of Telescopes, E38700, La Palma, Spain\\ 
$^{4}$ Instituto de Astrof\'{\i}sica de Canarias (IAC) and Universidad de La Laguna, Dpto. Astrof\'{\i}sica, Spain.\\
$^{5}$ Departamento de Astronom\'{\i}a, Universidad de Chile, Casilla 36-D, Santiago, Chile \\ 
$^{6}$ Department of Physics, University of Warwick, Gibbet Hill Road, Coventry, CV4 7AL, UK\\
}

\begin{document}

\date{Accepted XXXX. Received XXXX; in original form XXXX}

\pagerange{\pageref{firstpage}--\pageref{lastpage}} \pubyear{2016}

\maketitle

\label{firstpage}

\begin{abstract}

At the interface of \hii~regions and molecular gas peculiar structures appear,
some of them with pillar-like shapes. Understanding their origin is important for characterizing
triggered star formation and the impact of massive stars on the interstellar medium.
In order to study the molecular environment and the influence of the radiation on two pillar-like
features related to the \hii~region G46.5-0.2, we performed molecular line observations with the Atacama Submillimeter Telescope
Experiment, and spectroscopic optical observations with the Isaac Newton Telescope.
From the optical observations we identified the star that is exciting the \hii~region as a spectral type O4-6.
The molecular data allowed us to study the structure of the pillars and a \H~cloud lying between them.
In this \H~cloud, which have not any well defined \2 counterpart, we found direct evidence of star formation: two molecular 
outflows and two associated near-IR nebulosities. The outflows axis orientation is perpendicular 
to the direction of the radiation flow from the \hii~region. 
Several Class\,I sources are also embedded in this \H~cloud, showing that it is usual that the YSOs form large associations occupying 
a cavity bounded by pillars. On the other hand, it was confirmed that the RDI process is not occurring in 
one of the pillar tips.

\end{abstract}


\section{Introduction}

\hii~regions can generate dense layers of gas and dust in their surroundings. Very often, in the interface between the photo-dissociation
region (PDR) and the molecular gas peculiar structures can appear, some of them with pillar-like shapes and
cometary globules of dense gas. The pillars, also called ``elephant trunks'', usually present a column-like structure
with a head and a tail that physically connects with the molecular cloud. In our Galaxy they typically have a size between
1 and 4 pc, and a width between 0.1 and 0.7 pc \citep{gahm06}.
Most of the mass in these pillars is concentrated in their heads where, in some cases, signatures of star formation are found
\citep{ohlen13,smith10a,smith10b,sugitani89}.
The pillars were proposed to be transient features, being part of a continuous outwardly propagating wave of star formation 
driven by feedback from massive stars \citep{smith10b}. 
One of the main mechanism of star formation in such regions is the radiation-driven implosion (RDI),
first proposed by \citet{reip83}. This process
begins when the ionization front from the \hii~region moves over a molecular condensation such
as the head of a pillar-like feature, creating a dense outer shell of ionized gas, the ionized boundary
layer (IBL). If the IBL is over-pressured with respect to the molecular gas within the head of the pillar,
shocks are driven into it compressing the molecular material until the internal pressure is balanced with the pressure of
the IBL. At this stage the collapse of the clump begins a process leading to the creation of a new generation
of stars \citep{leflo94,berto90}. 
Recent theoretical and observational studies of pillars related to \hii~regions and possible star formation
linked to them can be found in \citet{hartigan15} and \citet{tremblin13,tremblin12}.
These studies highlight the importance of understanding the origin of these structures in order to shed light on issues such as
triggered star formation and the impact of massive stars on the initial mass function.

\begin{figure}
\centering
\includegraphics[width=8cm]{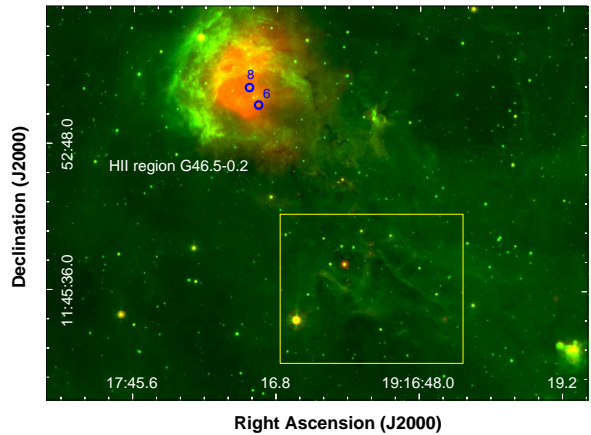}
\caption{Two-colour composite image of a large area towards the field of the \hii~region G46 where
the {\it Spitzer}-IRAC 8 $\mu$m emission is displayed in green
and the {\it Spitzer}-MIPSGAL emission 24 $\mu$m in red. The blue circles are the stellar candidates 
to be the exciting sources of the  \hii~region (sources 6 and 8 from \pap). 
Towards the southwest of G46, the region studied in this work, is delineated with a yellow rectangle.}
\label{present}
\end{figure}

G46.5-0.2 (hereafter G46) is an \hii~region located at a distance of about 4 kpc and spanning an angular size of 8\m\
(see \citealt{paron15} and references therein). It lies on the border of the molecular cloud GRSMC G046.34-00.21 \citep{rath09}.
In the south-west direction, about 10\m~away from G46 and associated with the molecular cloud
there are two pillar-like features that are bright at 8~$\mu$m and are oriented such that they point towards the open border of G46.
Figure \ref{present} shows a two-colour composite image of the G46 field
where the {\it Spitzer}-IRAC 8 $\mu$m emission is displayed in green and the {\it Spitzer}-MIPSGAL
at 24 $\mu$m in red. The region of the pillar-like features is shown with a yellow rectangle. 
Based on the pillars orientation and the similarity between the recombination line velocity of the \hii~region 
and the velocity of the molecular gas associated with the pillars, \citet{paron15} (hereafter Paper\,I) 
proposed that G46 and the pillar-like features are located at the same distance. 
Thus it was concluded that the pillars were carved and sculpted by the ionizing flux from G46.  
Through a photometric analysis presented in Paper\,I the authors look for candidates for the \hii~region exciting star(s). 
They find that sources 6 and 8 are the most likely to be O-type stars (blue circles in Fig.\,\ref{present}).
In addition, two concentrations of young stellar object (YSO) candidates were found and considered to be embedded in the molecular
cloud GRSMC G046.34-00.21. One of these YSO concentrations is closer to the G46 open border and consists of Class II type source candidates, 
and the other is mostly
composed of Class I type YSO candidates and is located just ahead of the pillar-like features, suggesting an age gradient in the YSO 
distribution. In an attempt to find a possible mechanism for star formation in the region, the authors performed
a rough pressure balance study using publicly available \3 J=1--0 data (46\s~in angular resolution) towards the tips of the pillars.
They found that the internal pressure of the neutral gas in the pillar-like features heads is larger than the external
pressure due to the ionized gas stalling at their tips, implying that RDI is not happening.

Taking into account that it is not common to find such large angular separation between \hii~regions
and their related pillars structures, and so well defined scenario without confusion from the usual clustering of 
\hii~regions as in G46, this is an interesting case to revisit it using better data in order to physically and chemically 
characterize the pillars and study the influence of the radiation on them. This new study points to
determinate whether the RDI process is ongoing or not, and to find direct evidence of star formation activity associated
with the pillars and their surroundings. Thus, using new molecular line observations
obtained with the Atacama Submillimeter Telescope Experiment (ASTE), optical spectroscopic observations from the 
IDS at the Isaac Newton Telescope (INT), and public near- and mid-IR data we performed a detailed study of the molecular 
environment towards the pillar-like features and the radiation flux arriving to them.

\section{Observations and data reduction}

\subsection{Optical observations}

In order to identify the source exciting G46 and obtain a more accurate estimate for the photon flux 
reaching the pillar tips, we performed spectroscopic optical observations of two stars, sources 6 and 8 from \pap~(blue 
circles in Fig.\,\ref{present}), which from photometric considerations were suspected to be O-type stars.
These observations were performed on 2016 May 12 with the Intermediate Dispersion Spectrograph (IDS) at the Isaac Newton Telescope 
on La Palma, Spain.
IDS observations were carried out using the EEV10 detector with the R632V grating (dispersion 0.9 \AA~pix$^{-1}$) 
in a central wavelength at 4800 \AA~providing a resolution 
of R $\sim$2500. Exposure times were 2400 sec and 300 sec for source 8 and source 6 respectively. The images were reduced and spectra 
extracted with IRAF software\footnote{IRAF is distributed by the National Optical Astronomy Observatory, which is operated 
by the Association of Universities for Research in Astronomy (AURA) under a cooperative agreement with the National Science Foundation.} 
following the standard procedures for long-slit optical spectroscopic observations. The spectra, in digital counts, were 
normalized to infer the spectral type of the stars.

\subsection{Molecular observations}

The molecular observations were performed on 2015 August 25-27  with the 10-m ASTE telescope \citep{ezawa04}. 
The CATS345 GHz band receiver, a two-single band SIS receiver remotely tunable in the LO frequency range of 324-372 GHz, was used. 
The XF digital spectrometer was set to a bandwidth and spectral resolution of 128 MHz and 125 kHz, respectively.
The spectral velocity resolution was 0.11 \ks~and the half-power beamwidth (HPBW) was 22\s~at 345 GHz. The system temperature
varied from T$_{\rm sys} = 150$ to 250 K. 
The absolute intensity calibration was made with observations of W51D (19:23:39.85, +14:31:10.1, J2000) 
and the intensity variation was estimated to be less than 7\%. The main beam efficiency was $\eta_{\rm mb} \sim 0.6$. 
The pointing accuracy was checked by observing RAql (19:06:22.254, +08:13:47.57, J2000) and it was within 5\s. 
The sky opacity varied from 0.04 to 0.06. 

\begin{figure}
\centering
\includegraphics[width=6.5cm]{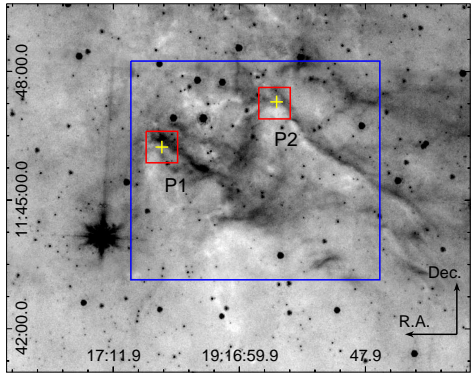}
\caption{{\it Spitzer}-IRAC 8 $\mu$m emission of the pillar-like features with the regions observed with ASTE: the larger 
blue box was mapped in the \2 J=3--2 and \H~J=4--3 lines, the small red boxes in the HCN J=4--3 line, and the crosses are
the positions of single pointings in the HNC J=4--3, \3 J=3--2 and CS J=7--6 lines.}
\label{obs}
\end{figure}

We observed a region of 5\farcm8$\times$5\farcm1~centered at 19:16:58.4, +11:45:41.6 (J2000) (blue box in Fig.\,\ref{obs}) with a
grid spacing of 22$^{\prime\prime}$~in the $^{12}$CO J=3--2 and HCO$^{+}$ J=4--3 transitions with an integration
time of 25 sec per pointing. As shown in Fig.\,\ref{obs} the
eastern pillar is called P1, while the western one is called P2. 
Additionally we observed two regions of $44^{\prime\prime} \times 44^{\prime\prime}$~with a grid spacing
of 22$^{\prime\prime}$~in the HCN J=4--3 line (red boxes in Fig.\,\ref{obs}). The centers of these regions are: 
19:16:56.6, +11:47:15.7 (J2000) (P1 head), and 19:17:07.3, +11:46:14.2 (J2000) (P2 head). The integration time was 500 sec per pointing. 
We also observed the heads of the pillar-like features with single pointings 
in the HNC J=4--3, \3 J=3--2 and CS J=7--6 lines  
with an integration time of 500 sec per pointing for the HNC and 120 sec per pointing for the other lines. The observed
positions, marked with yellow crosses in Fig.\,\ref{obs}, are: 19:16:56.4, +11:47:17.5 (J2000), and 19:17:07.3, +11:46:15.0 (J2000).
All observations were performed in position switching mode.

\begin{figure}
\centering
\includegraphics[width=9cm]{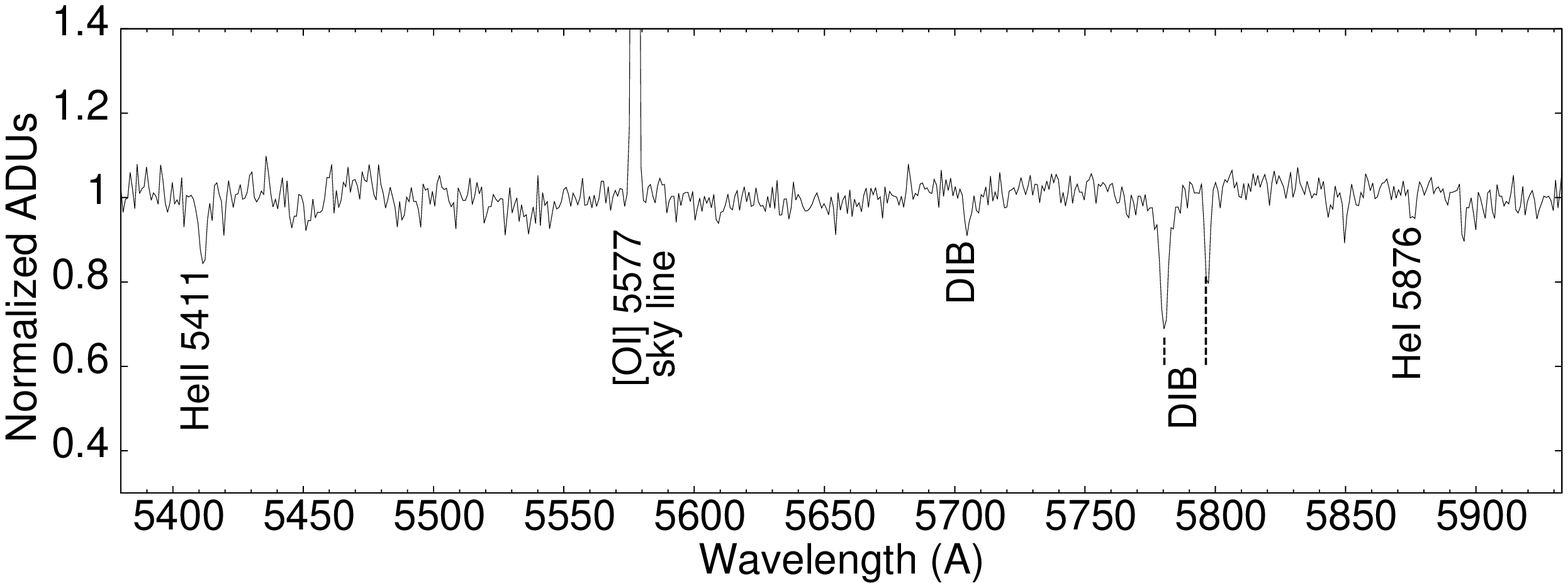}
\includegraphics[width=9cm]{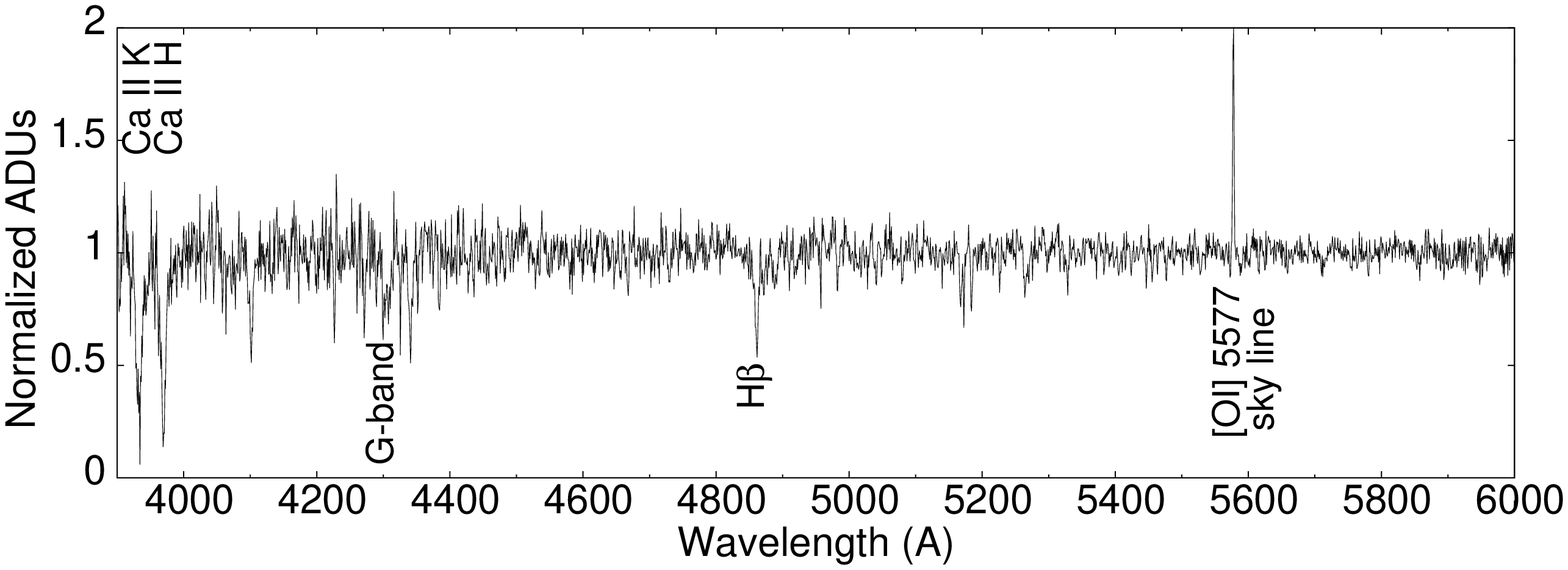}
\caption{Up: portion of the optical spectrum of source 8 where the absorption lines of He\,II $\lambda$5411 and He\,I $\lambda$5876 are present.
Some of the unlabeled absorption features are from diffuse interstellar bands (e.g. \citealt{lan15}). 
Bottom: optical spectrum of source 6, which following the Gray's Digital Spectral Classification Atlas it corresponds to a G-type star.}
\label{stars}
\end{figure}

\begin{figure}
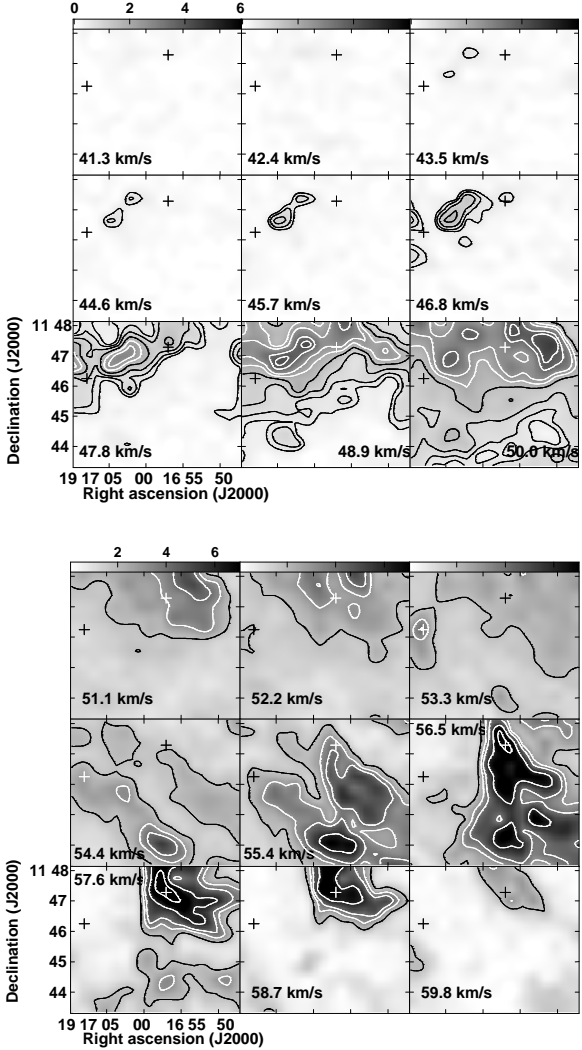

\centering
\includegraphics[width=7.2cm,angle=-90]{outflChan.ps}
\includegraphics[width=7cm,angle=-90]{pilaresChan.ps}
\caption{\2 J=3--2 velocity integrated channel maps each 1.1 \ks~of the surveyed region are shown
in two images. From 41.3 to 50.0 \ks~with contours levels of 0.5, 0.7, 1.0, 1.5, 2.0, 2.5, 3.5, and 4.0 K \ks~(upper image), 
and from 51.1 to 59.8 \ks~1.8, 3.0, 4.0, 6.0, and 8.0 K \ks~(bottom image). The intensity range is shown at top of each 
image and is in K \ks. The crosses are the positions of the pillars heads.}
\label{channA}
\end{figure}

\begin{figure*}
\centering
\includegraphics[width=17cm]{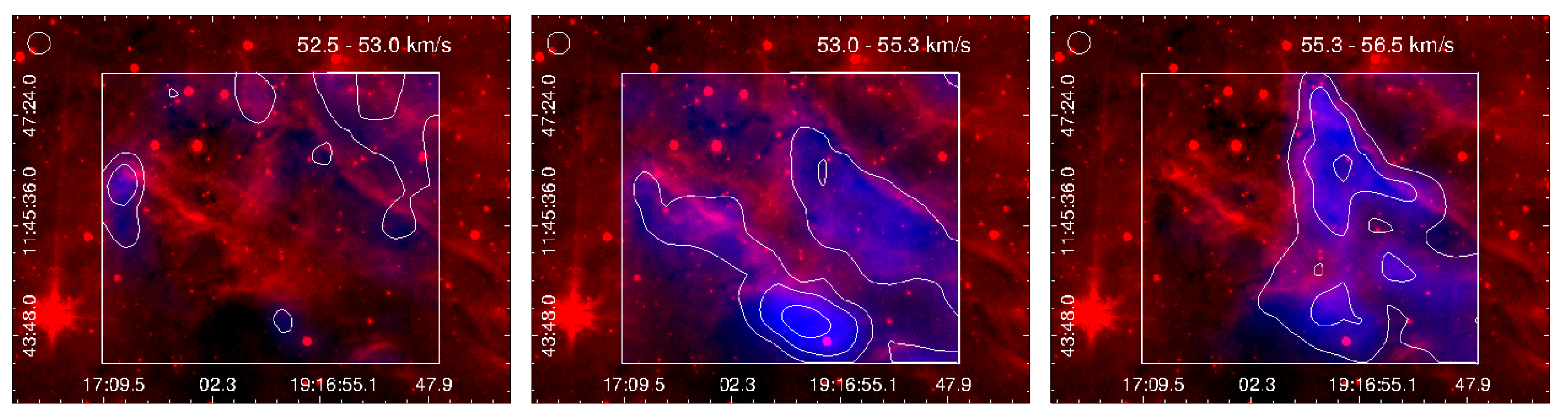}
\caption{\2 J=3--2 velocity integrated maps (blue with white contours) displayed over the
IRAC-Spitzer 8 $\mu$m emission (red). The integration velocity interval is shown at the top
of each panel and the beam of the molecular observations is included to the top left corner.
The contours levels are:  0.7 and 0.9 K \ks~(left), 3.6, 5.0, and 7.0 K \ks~(middle),
and 3.0 and 4.9 K \ks~(right).}
\label{pilarsCO}
\end{figure*}

The data were reduced with NEWSTAR\footnote{Reduction software based on AIPS developed at NRAO,
extended to treat single dish data with a graphical user interface (GUI).} and the spectra processed using the XSpec software
package\footnote{XSpec is a spectral line reduction package for astronomy which has been developed by Per Bergman at Onsala
Space Observatory.}.
The spectra were Hanning smoothed to improve the signal-to-noise ratio, and in some cases a boxcar smoothing was also applied.
Polynomials between first and third order were used for baseline fitting.

\section{Results}

\subsection{Identifying the stars causing G46 excitation}

From the optical spectroscopic analysis of sources 6 and 8 (see Fig.\,\ref{present}) we 
determine that source 6 cannot be the contributor
to the ionization of the region. Following the Gray's Digital Spectral Classification 
Atlas\footnote{http://ned.ipac.caltech.edu/level5/Gray/frames.html}, its spectrum (Fig.\,\ref{stars} bottom) suggests 
that this source is a G-type star, which would be unable to generate an \hii~region. 
Source 8 suffers high-extinction (B--R$\sim$3.6) being relatively faint at blue wavelengths, hence the blue portion of the spectrum, 
which contains the He lines commonly used for O spectral classification is very noisy. Nevertheless, in the range from 5400 to 5900 \AA~the 
lines of He\,II $\lambda$5411 and He\,I $\lambda$5876 are clearly visible (see Fig.\,\ref{stars} up). 
The ratio [He\,II]/[He\,I] from these two lines corresponds well with 
the spectral-type sequence, in which, according to \citealt{walb80}, He\,I $\lambda$5876 is present from type O4 to later types, 
and a ratio $\sim1$ corresponds to a type O7. 
Therefore, according to the observed ratio [He\,II $\lambda$5411]/[He\,I $\lambda$5876] for source 8 and comparing with ratios of 
[He\,II $\lambda$4541]/[He\,I $\lambda$4471] 
and [He\,II $\lambda$4541]/[He\,I $\lambda$4387] in \citealt{walb90}, we can infer that source 8 is a O4-6 star.

\subsection{Molecular environment related to the pillars}

As studied in Paper\,I and mention here in Sect.\,1, the velocities of the molecular gas concentrations related to the pillars 
are similar to the recombination line velocity of the \hii~region G46, thus along this work it is assumed that
all of these structures are located at the same distance.

We analysed the \2 J=3--2 data cube along the whole velocity axis in order to find molecular structures
associated with the pillar-like features delineated by the IR emission at 8 $\mu$m.  
In Fig.\,\ref{channA} the \2 integrated emission is presented in channel maps in steps of 1.1 \ks~along
the velocity interval from 41 to 60 \ks. The positions of the pillars heads are indicated with crosses in
each panel. Abundant molecular gas towards the north of the surveyed region and not
related to the pillars can be distinguished in the range 43--53 \ks.  Between 43.5 and 48.9 \ks~there is an
interesting structure with a bilobed shape towards the northeast and lying between both pillars, which will be studied in 
Sect.\,\ref{outflsect}. 
It can be appreciated that in the velocity interval of 53.3--56.5 \ks~there are some molecular 
structures that are likely associated with the pillar-like features, while in the other velocities
it appears clumpy molecular structures without any morphological correspondence with the pillars. 

The molecular structures associated with the pillars extend along narrow velocity intervals. This is shown in 
Fig.\,\ref{pilarsCO} where the molecular gas emission is superimposed to the 8 $\mu$m emission. 
The most remarkable morphological association between the molecular gas and the 8 $\mu$m emission is between 55.3 and 56.5 \ks~in 
the pillar P2 (right panel in the Figure). Then, between 53.0 and 55.3 \ks~(middle panel), there is a weak feature related 
to P2, and another one associated with P1 with a peak towards its base. A clump associated with the P1 tip appears between 
52.5 and 53.0 \ks~(left panel), showing that the molecular gas associated with this pillar is                  
more clumpy than the gas associated with P2.

\begin{figure}
\centering
\includegraphics[width=5.8cm]{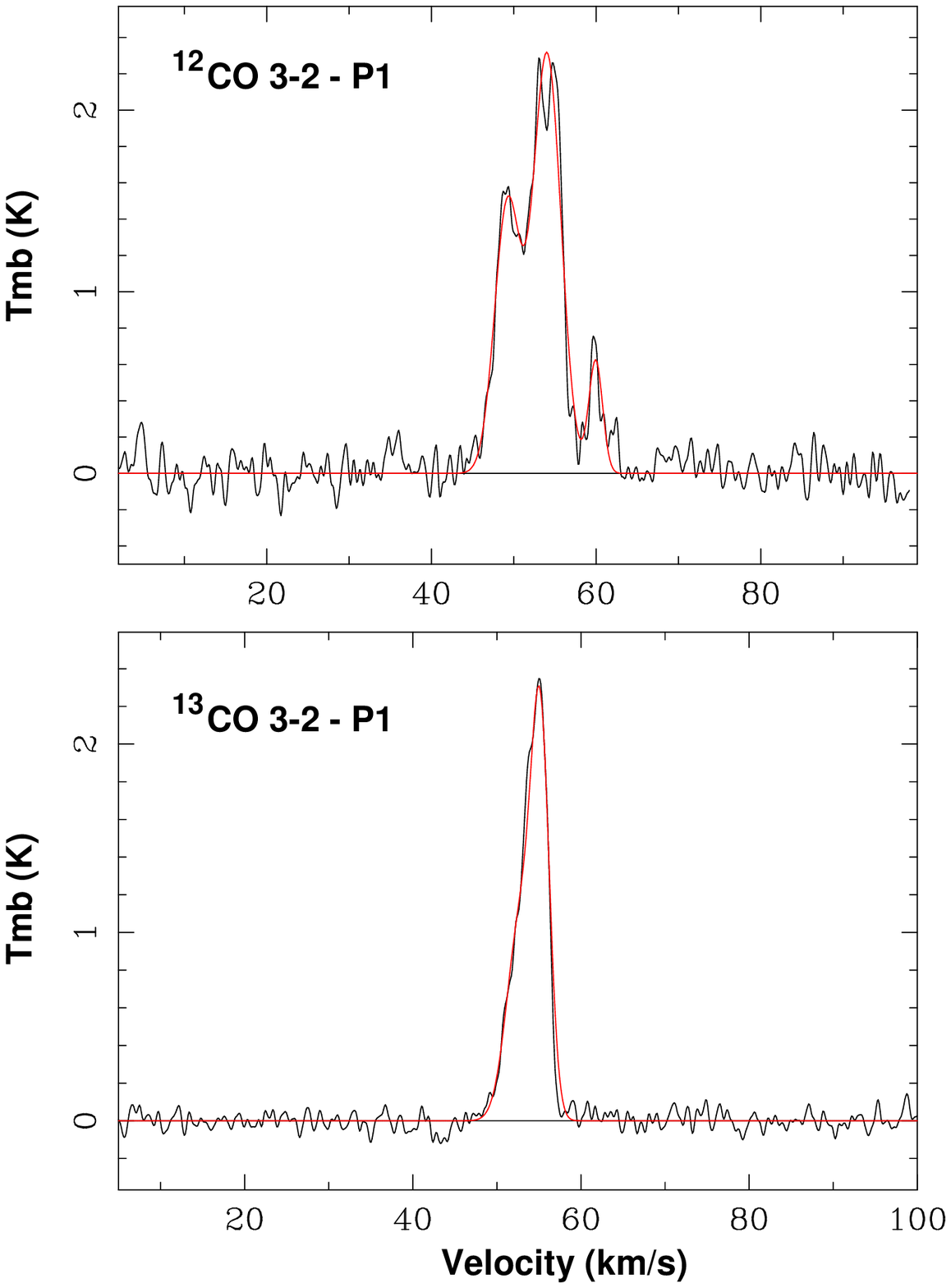}
\includegraphics[width=5.8cm]{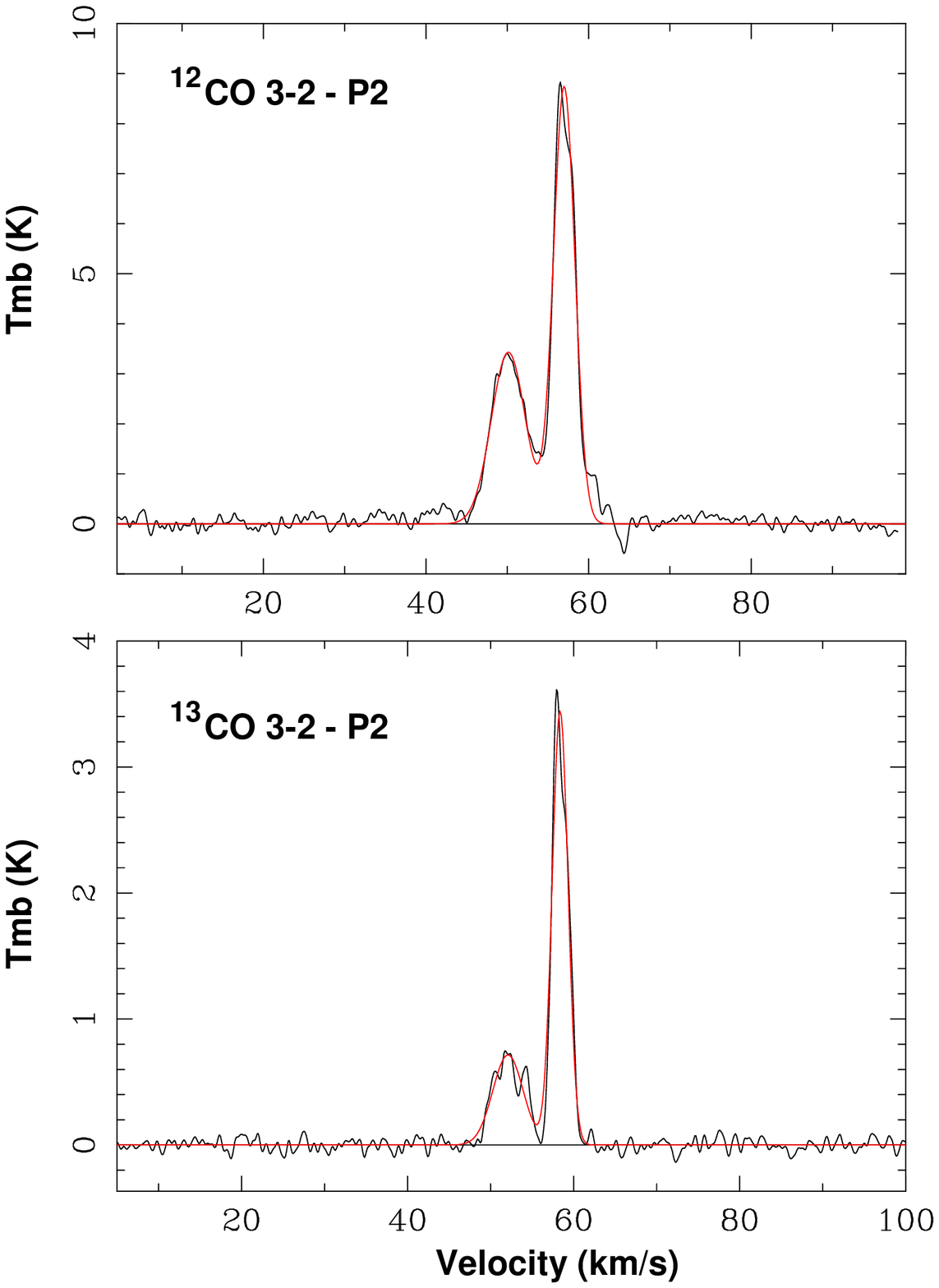}
\caption{\2 and \3 spectra observed in the region of the heads of the pillar-like features (yellow
crosses in Fig.\,\ref{obs}). The Gaussian fitting to each spectrum is shown in red. The rms noise is about 100 and 50 mk for the \2 and \3
lines, respectively. }
\label{spectco}
\end{figure}

Figure\,\ref{spectco} shows the \2 and \3 spectra obtained in the location of the yellow crosses shown in Fig.\,\ref{obs}, i.e. the heads
of the pillar-like features. The CS J=7--6 emission was not detected at these positions. 
Table\,\ref{paramco} lists the line parameters obtained from Gaussian fits to these spectra
(superimposed in red in the figure). The \2 spectrum from the head of P1 was fitted with three Gaussians. Two
main components within the velocity range 45--55 \ks~can be seen with a smaller component centered at $\sim60$ \ks. The \3 spectrum is 
less complex and could be fitted with two Gaussians, the counterparts of the \2 main components.
In the case of the P2 head, both CO isotopes were fitted with two Gaussians showing the presence of two molecular components
towards this region.

\begin{table}
\footnotesize
\caption{Line parameters for the CO isotopes.}
\label{paramco}
\begin{tabular}{lcccc}
\hline
Line              & T$_{\rm mb}$  & v$_{\rm LSR}$    & $\Delta$v    \\
                  &   (K)         &   (\ks)          &  (\ks)             \\
\hline
\2 (3--2) - P1      &  1.48         & 49.23            &  3.70       \\
                  &  2.30         & 54.03            &  5.00      \\
                  &  0.62         & 59.94            &  1.97       \\
\3 (3--2) - P1      &  1.06         & 52.82            &  3.83       \\
                  &  1.91         & 55.20            &  2.53       \\
\hline
\2 (3--2) - P2      &  3.43         & 50.13            &  4.97       \\
                  &  8.73         & 57.00            &  3.10       \\
\3 (3--2) - P2    &  0.71         & 52.10            &  4.30       \\
                  &  3.44         & 58.36            &  2.24       \\
\hline
\end{tabular}
\end{table}

To derive the molecular mass contained in the heads of the pillar-like features, local thermodynamic equilibrium (LTE) was assumed.
Following the standard LTE procedures we obtain a \2 column density, N(\2), for each pillar head. 
For the details of the formulae see e.g. \citet{paron12}.
{\bf From the ratio of the $T_{mb}$(\2) to $T_{mb}$(\3) peaks can be derived the $^{12}$CO and $^{13}$CO optical depths ($\tau_{12}$, 
$\tau_{13}$). 
As shown in Fig.\,\ref{spectco} and in Table\,\ref{paramco}, the $^{12}$CO and $^{13}$CO spectra have more than one component, 
and thus at least a couple of $\tau_{12}$ 
and $\tau_{13}$ could be obtained for each pillar head. By inspecting the central velocities and $\Delta$v of each component 
(see Table\,\ref{paramco})  we selected 
the $T_{mb}$(\2) and $T_{mb}$(\3) peaks that are within the velocity intervals in which each pillar extends (see Fig.\,\ref{channA}), thus 
we used the components at $\sim$55 \ks~and $\sim$58 \ks~for P1 and P2, respectively.
Assuming a canonical [\2]/[\3] isotope abundance ratio of 50, we estimate the optical depths in 
$\tau_{12} \sim 89$ and $\tau_{13} \sim 1.7$ for the P1 head, and $\tau_{12} \sim 25$ and $\tau_{13} \sim 0.5$ for the P2 head.
}
Due to the high optical depth of the \2 emission, it is not possible to obtain reliable mass values from the \2 column density. Thus
using the derived $\tau_{13}$ values we calculate the \3 column densities from each single pointing at P1 and P2. The excitation
temperature (T$_{\rm ex} \sim 10$ K used in both cases) was obtained from the \2 spectra at the tips of each pillars.  
Assuming that the \3 emission is uniformly distributed in the pillars heads, i.e. it is assumed that the column density value is the
same at all beam positions within a circle delimited by the curvature of the tip of the pillars, we estimate a total \3 column density 
for each pillar head.  
Finally, the H$_{2}$ column densities were derived from N(H$_{2}$) $=$ N(\3)/$X_{^{13}\rm{CO}}$, where $X_{^{13}\rm{CO}} = 2 \times 10^{-6}$ 
(e.g. \citealt{yama99}), and the mass 
is derived from:
\begin{equation}
        {\rm M}  = \mu m_{H} D^{2} \Omega {\rm N(H_{2})}
\label{eqM}
\end{equation}
\noindent
where $\Omega$ is the solid angle subtended by the beam size, $D$ is the distance of 4 kpc, $m_{H}$ is the hydrogen mass, 
and $\mu$ is the mean molecular weight, assumed to be 2.8 by taking into account a relative helium abundance of 25\%. 
The obtained masses for the head of P1 and P2 are about 60 \msun~and 30 \msun, respectively.

\begin{figure}
\centering
\includegraphics[width=8cm]{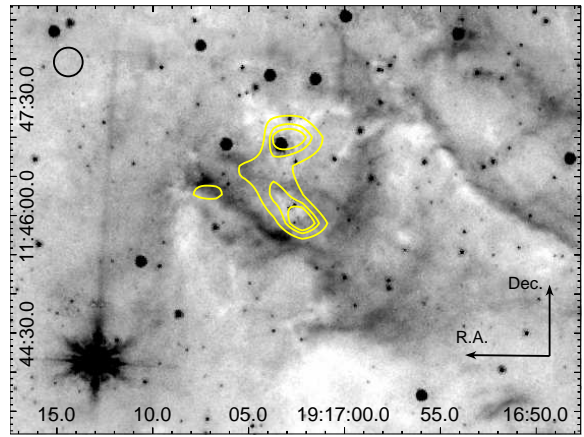}
\caption{The \H~J=4--3 emission integrated between 48 and 60 \ks~is displayed in contours
over the {\it Spitzer}-IRAC 8 $\mu$m emission. The contours levels are 0.5, 0.6, and 0.7 K \ks.
The beam of the molecular observation is included in the top left corner.}
\label{hco+map}
\end{figure}

\begin{figure}
\centering
\includegraphics[width=6cm]{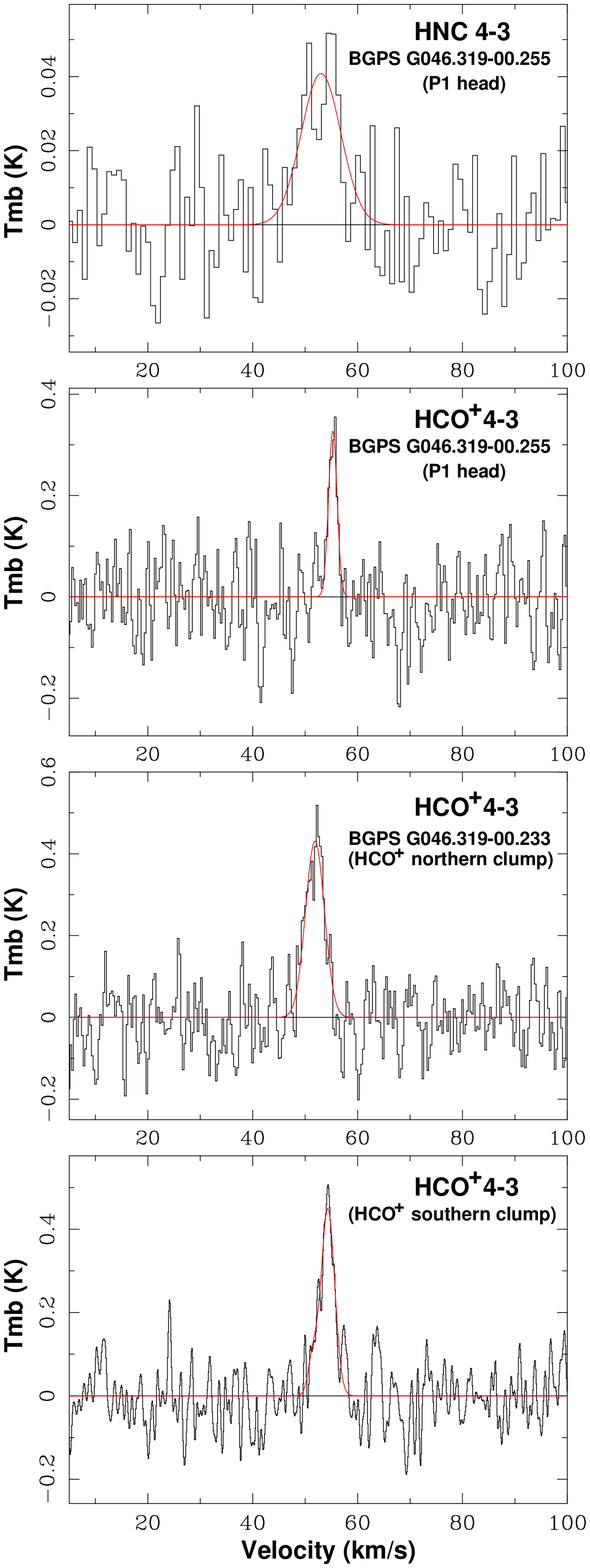}
\caption{First and second panels: HNC and \H~J=4--3 spectra obtained in the region of P1 head, at the position of BGPS G046.319-00.255. 
Third and fourth panels: \H~J=4--3 spectra at the peak position of BGPS G046.319-00.233 at the northern \H~clump and
at the southern \H~clump, respectively (see Fig.\,\ref{hco+map}). A Gaussian fit of each spectrum is shown in red. 
The rms noise for these spectra are 13 and 45 mk for the HNC and the \H~lines, respectively.  }
\label{spectraH}
\end{figure}

Analyzing the \H~J=4--3 cube we found emission in a region between the 
pillar-like features (see Fig.\,\ref{hco+map}). The \H~structure is composed of two clumps (hereafter \H~northern and southern clumps).
The centre of the northern one coincides in projection with the peak of the 1.1 mm emission source BGPS G046.319-00.233 \citep{roso10}
and with the \2 bilobed structure described above. Besides this positional coincidence, it is worth noting that at the \H~structure
position, i.e. the region between both pillars, there is not a well defined \2 structure (see Fig.\,\ref{channA}).
In addition, some weak \H~emission appears towards the head of P1 
in coincidence with the source BGPS G046.319-00.255. At this position we also detected HNC J=4--3, which is shown in
Fig.\,\ref{spectraH} together with the \H~spectra towards the peak of the mentioned BGPS sources and the southern \H~clump. 
The line parameters obtained from Gaussian fits to these spectra are presented in Table\,\ref{param}. 
Towards the head of P2, which also coincides with a millimeter continuum source (BGPS G046.314-00.213), was not detected
\H J=4--3. Finally, HCN J=4--3 emission was not detected in either of the pillar-like features.

With the parameters obtained from our data and those for the \H~J=3--2 from \citet{shirley13}  
also included in Table\,\ref{param}, we performed a non-LTE study 
of P1 head and the northern \H~clump using the Radex code \citep{vander07}. 
In order to compare both sets of data, our \H~J=4--3 spectra
were convolved with a beam of 30\s~from the Shirley et al. \H~J=3--2 observations.
The Radex calculations were made for different kinetic temperatures in the range between 20
and 100 K. The results are presented  
in Table\,\ref{tradex}. In the case of P1 head Radex does not converge for a T$_{\rm k} = 100$ K.

\begin{table}
\footnotesize
\caption{Line parameters for the \H~and HNC lines.}
\label{param}
\begin{tabular}{lcccc}
\hline
Line    & T$_{\rm mb}$  & v$_{\rm LSR}$    & $\Delta$v  & $\int{\rm T_{mb} dv}$  \\
                  &   (K)             &   (\ks)          &  (\ks)            &           (K \ks)      \\
\hline
\multicolumn{5}{c}{ BGPS G046.319-00.233 (\H~northern clump)  }\\
\hline
\H (4--3)        &   0.43              &  51.87         &  4.25             &   1.87          \\
\H (3--2){\bf *}       &   0.53              &  52.40         &  3.80       &  2.27           \\
\hline
\multicolumn{5}{c}{(\H~southern clump)  }\\
\hline
\H (4--3)        &   0.45              &  54.32         &  3.05             &   1.68          \\
\hline
\multicolumn{5}{c}{ BGPS G046.319-00.255 (P1 head) }\\
\hline
\H (4--3)        &   0.33              &  55.26         &  1.90             &     0.68        \\
\H (3--2){\bf *}       &   0.35              &  54.70         &  3.20       & 1.03           \\
HNC (4--3) &   0.04             &  53.02         &  8.90          &  0.37             \\
\hline
\multicolumn{5}{l}{\footnotesize {\bf *} from \citet{shirley13}.} \\
\end{tabular}
\end{table}

\begin{table}
\centering
\caption{Radex Results}
\label{tradex}
\begin{tabular}{lcc}
\hline
T$_{\rm k}$ (K)    & N (cm$^{-2}$)  & n (cm$^{-3}$)    \\
\hline
\multicolumn{3}{c}{ BGPS G046.319-00.233 (\H~northern clump)  }\\
\hline
20   & $9.7\times10^{11}$  & $4.9\times10^{6}$                                \\
30   & $9.3\times10^{11}$  & $1.7\times10^{6}$                  \\
50   & $9.6\times10^{11}$  & $7.3\times10^{5}$                   \\
100  & $1.1\times10^{12}$  & $3.2\times10^{5}$                   \\
\hline
\multicolumn{3}{c}{ BGPS G046.319-00.255 (P1 head) }\\
\hline
20   & $7.1\times10^{11}$  & $1.4\times10^{6}$                 \\ 
30   & $8.5\times10^{11}$  & $5.3\times10^{5}$                 \\ 
50   & $1.7\times10^{12}$  & $1.2\times10^{5}$                   \\
100  & --                  & --                   \\
\hline
\end{tabular}
\end{table}

Additionally we estimate the dust temperature, $T_{d}$, and the H$_{2}$ column density for P1 head and the \H~northern clump from 
the {\it Herschel} public data at 160, 250, 350, and 500 $\mu$m (OBsId 1342207054 and 1342207055) and the ATLASGAL data at 870 $\mu$m 
(from \citealt{urqu14}). 
Assuming dust emission in the optically thin regime the surface brightness, I$_\nu$, can be expressed as a gray-body 
function for a single temperature
\begin{equation}
        I_{\nu} = \kappa_{\nu}(\nu/\nu_{0})^{\beta} B_{\nu}(T_{dust}) \mu  m_{H} {\rm N(H_{2})}
\label{eqD}
\end{equation}
\noindent
where $B_{\nu}(T_{dust})$ is the blackbody function for a dust temperature $T_{dust}$, $\mu$ 
the conversion factor from H$_{2}$ to total gas mass,
assumed to be 2.8 by considering an He abundance of 25\%, 
$m_{H}$ the atomic hydrogen mass and $\kappa_{\nu}(\nu/\nu_{0})^{\beta}$ 
the dust opacity per unit mass, where $\nu_{0}$ is assumed to be 0.1 cm$^{2}$g$^{-1}$ 
at 1 THz \citep{beck90} under a gas-to-dust ratio of 100 and $\beta=2$ \citep{anderson12}. 
Then, we performed an SED fitting with $T_{dust}$ and N(H$_{2}$) as free parameters
where the surface brightness was obtained from the level 2.5 {\it Herschel} data and from
the catalogued ATLASGAL sources. All the emissions were convolved to the same
resolution 44\s, and rebinned to the same pixel size 14\s. We determined that $T_{d}$ and N(H$_{2}$) 
are about 18 K and $1.8\times10^{22}$ cm$^{-2}$, 
and 20 K and $0.8\times10^{22}$ cm$^{-2}$, for the \H~northern clump and the P1 head, respectively.
The errors involved in the temperatures and column densities are about 10\% and 40\%, respectively.
From these results, if we assume that the dust and gas are coupled, and thus $T_{d} = {\rm T_{k}}$,
we can favour the Radex results for T$_{\rm k} = 20$ K for both molecular concentrations.
In this way we estimate the \H~abundances,
$X_{\rm HCO^{+}} \sim 5.4 \times 10^{-11}$ and $\sim 8.8 \times 10^{-11}$ for the \H~northern clump and P1 head, respectively.

Additionally from the same SED fitting procedure it was obtained the $T_{dust}$ and N(H$_{2}$) for the P2 head
(source BGPS G046.314-00.213) in about 18 K and $1\times10^{22}$ cm$^{-2}$. Using the N(H$_{2}$) obtained in P1 and P2 
from the SED fitting in Eq.\,\ref{eqM} we estimate the masses in an independent way as done above with the \3 emission, 
obtaining about 80 \msun~for both pillars head.

\subsection{Direct evidence of star formation in the region}
\label{outflsect}

\begin{figure}
\centering
\includegraphics[width=8cm]{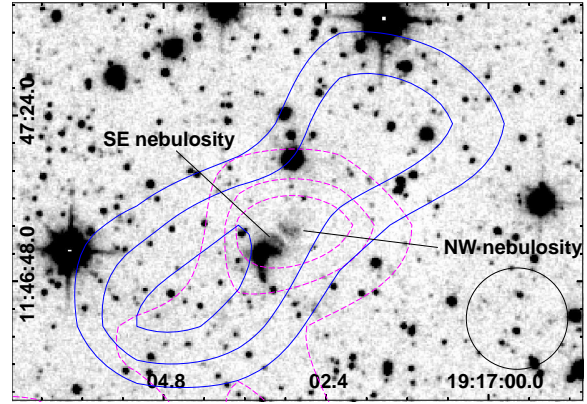}
\caption{{\it Ks}-band emission obtained from the UKIDSS Survey with blue contours of the \2 J=3--2 emission integrated
between 42 and 50 \ks~with levels of 6.0, 7.0, and 8.5 K \ks. The dashed magenta contours are the same \H~contours as presented in 
Fig.\,\ref{hco+map}. The beam of the molecular observation is 
included in the bottom right corner.}
\label{outflows}
\end{figure}

Given that in \pap~it was shown the existence of several YSO candidates in the region between the pillars, 
we inspected the {\it Ks}\,-band emission obtained from the UKIDSS Survey looking for signatures
of extended emission likely to be related to outflow activity. We found an interesting structure composed of
two nebulosities extending from southeast to northwest lying at the \H~northern clump related to the BGPS G046.319-00.233 source 
(see Fig.\,\ref{outflows}). The figure also demonstrates that both nebulosities are separated by a region
of low {\it Ks} emission. Interestingly, this near-IR structure lies at the same position of the northeastern bilobed molecular feature 
that appears between 43.5 and 48.9 \ks~shown in Fig.\,\ref{channA}.
By integrating the \2~emission along the  velocity range 42--50 \ks, two conspicuous molecular lobes appear 
(blue contours in  Fig.\,\ref{outflows}) that extend, as the near-IR features, from southeast to northwest.  
The near-IR features mentioned above are located, in projection, almost at the center of these lobes, 
which strongly suggest that we are observing molecular outflows generated by the same source that produced the near-IR features.
The lengths of the outflows are 50\s~and 44\s, for the northwestern and southeastern lobes, respectively, which at the distance
of 4 kpc implies about 0.9 and 0.8 pc, respectively.

To estimate the outflow mass, following \citet{bertsch93}, we calculated the H$_{2}$ column density from
\begin{equation}
{\rm N(H_{2})} = 2.0 \times 10^{20} ~\frac{\rm W(^{12}CO)}{\rm [K~km~s^{-1}]}~ {\rm (cm^{-2})}, 
\label{eqMb}
\end{equation}
\noindent
where W(\2) is the \2 J=3--2 integrated intensity.
The mass was derived from Eq.\,\ref{eqM}, obtaining 10.5 and 9.0 \msun~for the southeastern
and northwestern lobes respectively.

\section{Discussion}

From the new molecular data presented in this work we confirm that the gas related to the 
pillar-like features embedded in the large molecular cloud GRSMC G046.34-00.21 and associated with the \hii~region G46 extends 
along narrow velocity intervals. These pillars contain embedded cold submm cores as was found toward similar structures in 
the Eagle Nebula \citep{white99}.
The derived masses for P1 and P2 heads from the new molecular observations are about 60 and 30 \msun, respectively.
These values are in quite agreement with the mass obtained from the dust emission (about 80 \msun~in both pillar heads).
It is worth noting that these values are one order of magnitude lower than those obtained in \pap~using the \3 J=1--0 line. 
Given that this discrepancy is large and it can not be explained just through the use of different lines, the values from 
\pap~were revised. After a carefully inspection of the mass estimate procedure done in \pap~we noticed that it was wrongly used a larger 
area than the actual extension of the pillars heads, yielding overestimated values. The actual mass values derived from the \3 J=1--0 line are
between 60 and 80 \msun~for both pillars heads in completely agreement with the values obtained here. It is important to remark
that the volumetric density values in \pap~do not change significantly, and hence the RDI analysis and its conclusion do not change.

We conclude that the mass of the pillars heads ranges from $\sim$30 to $\sim$80 \msun, which 
is in agreement with masses of pillars heads found in several regions \citep{gahm06,white99} and with values obtained from 3D dynamical 
models used to study the formation of these kind of structures \citep{mackey10}.

In this work we have mainly analysed three molecular structures: 
the pillar-like features P1 and P2, and a cloud observed in the \H~J=4--3 line lying between both pillars and composed of 
two clumps. Even though these features are embedded in the 
same large molecular cloud, we found different physical conditions between them. We suggest that it could be due
to the fact that these features may be exposed in different ways to the UV radiation. 
It is possible that P2 is more exposed to the radiation flow from G46 than P1. 
The non-detection of \H~towards P2, and the \H~abundances obtained
in P1 and the \H~northern clump may be explained through different ionization rates within the region. 
As is stated in \citet{goico09},
\H~is mainly destroyed via dissociative recombination, and thus its abundance is inversely proportional to the electron abundance. 
The structures belonging to the large and clumpy molecular cloud that are more exposed to the radiation from the \hii~region 
are expected to have a higher electron abundance than the less exposed ones. We remark that this is quite a simple 
conclusion based only in the \H~analysis, and as \citet{goico09} pointed out, it is necessary to consider also the abundances of metals, 
PAHs, other ions, and cosmic rays. In the case of the \H~northern clump this situation is more complex because, as it is
presented and discussed in Sects.\,\ref{outflsect} and \ref{sectOutfl}, this clump presents outflow activity, which according 
to \citet{rawlings04} may increase the \H~abundance.
Nevertheless our result gives support to different ionization rates along the region.

From the optical spectroscopic results of the star named source 8 (an O4-6 type star), together with the previous photometric
analysis (see \pap) and considering its location, exactly at the center of G46 \hii~region, we can conclude that this source is indeed
responsible for the ionization of the region. This more accurate determination of the spectral type of the ionizing star allows us 
to derive a better estimate of the amount of UV photons arriving at the pillars.
Given that P2 could be the more exposed structure to the radiation from the \hii~region, 
the analysis was done only for this structure. Following the same procedure as described in \pap~we estimate an upper limit 
for the predicted photon flux impinging on the P2 head in
$\phi_{pred} \sim 3 \times 10^8$ cm$^{-2}$s$^{-1}$. Thus, according to \citet{bis11}, who suggested that triggered 
star formation through RDI occurs 
only when $10^9$ cm$^{-2}$s$^{-1} \leq \phi \leq 3 \times 10^{11}$ cm$^{-2}$s$^{-1}$, 
we can conclude that in the head of P2 it is unlikely that triggered star formation is ongoing.

\subsection{HCN and HNC in the pillars}

The HCN J=4--3 line was not detected in any pillar and the 
HNC J=4--3 line was detected only at P1 tip. 
Taking into account that usually both isomeres are detected in the same region and it is expected that the HNC/HCN ratio is on 
the order of unity 
in cold regions \citep{herbst78,schil92},
it is intriguing to consider why at the P1 tip, a region that coincides with a catalogued dark cloud, HNC is detected and not HCN. 
It could be that the HNC has higher abundance than the HCN, being the HCN abundance low enough to be below the detection 
limit for these observations. This is indeed possible because, according
to \citet{hirota98}, the recombination reaction HCNH$^{+}$ + e$^{-}$ $\rightarrow$ HNC + H, HCN + H, one of the main possible chemical paths 
for the formation of these isomeres, leads a higher abundance of HNC than of HCN. On the other hand, 
\citet{chenel16} point out that in the interstellar regions exposed to an intense UV radiation field, HCN should be
more abundant than HNC. The destruction of both isomers in such 
regions is dominated by photodissociation rather than by chemical reactions with radicals or 
ions, and the authors found that HNC is destroyed faster than HCN. In the case of the P1 head, it is likely that this structure is not
strongly irradiated by the UV photons and hence the UV destruction 
mechanism is not ongoing.  
At the P2 head, which may be more exposed to the radiation, it could
be possible that the HCN and HNC have been totally destroyed by the UV radiation.
This is in agreement with the results from the \H analysis discussed in the previous Section.

\subsection{Confirming star formation}
\label{sectOutfl}

We find direct evidence of star formation at the \H~northern clump in both near-IR and molecular line emissions.
The near-IR emission shows the presence of two nebulosities separated by a region of low emission. 
Taking into account that the discovered molecular outflows extend along the same direction as the near-IR nebulosities, 
we suggest that the IR emission arises from cavities cleared in the
circumstellar material. These kind of cavities can be generated by the action of winds from the YSO \citep{shu95,reip01}, 
or by precessing jets that clear the circumstellar material (e.g. \citealt{kraus06}).  
The orientation of the nebulosities and the analysis of the molecular lobes strongly suggests that the outflows extend
mainly in the plane of the sky. This may explain why a central point source is not seen at near-IR emission,
it being probable that we are observing the YSO system in an edge-on orientation and the central source is likely hidden behind the disc 
(e.g. \citealt{perrin06}).
In this scenario the central source and the disc must lie in the region of low near-IR emission between the two 
nebulosities. This YSO system and most of the Clase\,I YSO candidates found in \pap~are embedded in the mapped \H~cloud 
lying between both pillars, which based on what was observed by \citet{smith10a} in the Carina Nebula, seems to be common. As 
the authors concluded, the YSOs tend to form large associations occupying a cavity that is bounded by pillars. It is important
to note that the observed \H~cloud has not a well defined counterpart in the \2 emission, suggesting that the external layers of
the molecular gas traced by the \2 emission were disrupted by the massive star feedback. 

The mass and size of the discovered molecular outflows are similar to those found at several sources catalogued as massive
molecular outflows \citep{beuther02}, suggesting that we are indeed observing a massive YSO.
It is worth noting that its outflows axis orientation is perpendicular to the direction of the \hii~region
open border, and hence to the direction of the radiation flow. This orientation was also found in several YSOs in the Carina 
Nebula \citep{smith10a}. Using 3D simulations, \citet{lora09} proposed that the interaction between an advancing ionization front
and a neutral cloud, wherein compressed clumps of gas that form at the unstable interface tend to collapse and have angular momentum
vectors perpendicular to the direction of the radiation flow. The YSO system described here could be considered as a new observational 
evidence of this phenomena, aiming for further theoretical studies in which the large distance (about 10 pc) between the YSO and the
radiation source should be taken into account.

\section{Summary and concluding remarks}

Using the ASTE telescope and public IR data we investigated in detail the 
molecular environment related to two pillar-like features likely generated by the 
\hii~region G46.5-0.2 (G46). Additionally, using the INT telescope we found the star 
that is exciting this \hii~region. The main results of this study are summarized as follows:

(1) From the \2 J=3--2 emission we found that the molecular structure of the pillar-like features 
extend along narrow velocity intervals. It was observed that P2 has the expected morphology of the pillars, 
while the molecular emission at P1 is more clumpy and the main mass concentration is not in its head.
P1 may be in an earlier evolutionary stage in the formation of these kind of pillars.

(2) The \H~J=4--3 emission was found to be concentrated in a cloud lying between both pillars and composed of two
clumps. This cloud has not any well defined \2 counterpart. Faint \H~emission was detected towards the head 
of P1 where HNC J=4--3 emission was also detected. At the P2 tip, emission from these species was not detected. 
From a \H~abundance analysis it is suggested that P2 could be more exposed to the radiation from G46 than P1. 

(3) From the optical spectroscopic observation it was determined that the so-called source 8 (from \pap)
is an O4-6 star. Thus, from its position and spectral type we conclude that this is the source responsible
for ionizing the region.

(4) From the spectral type determination of the ionizing source we obtained a more accurate value for 
the amount of UV flux arriving at P2. We conclude that it is unlikely that the 
RDI process is ongoing in the P2 head.

(5) We found direct evidence of star formation towards the cloud mapped in the \H~emission. 
From the \2 J=3--2 line we detected two massive molecular outflows extended mainly along the plane of the sky 
coinciding with two nebulosities separated by a region of low 
emission seen at near-IR emission. We propose that we are observing an YSO system in an edge-on position.

The confirmed star formation activity in the analysed region is occurring in a molecular clump that it is bounded by 
the pillar-like features, which it seems to be common to find in such regions, confirming that the pillars
are transient structures that are part of a continuous outwardly propagating wave of star formation
driven by the feedback from massive stars as proposed by \citet{smith10b}. 
The mapped \H~cloud, in which it is embedded the discovered outflows source and most of the Class I sources shown in Paper\,I, 
has not a well defined counterpart in the \2 emission. It suggests that the external layers of the molecular cloud 
were disrupted by the radiation from G46, which is in agreement with a scenario of star formation
driven by the radiation from massive stars. 
Additionally, it was found that the outflows axis orientation of the discovered YSO is perpendicular to the direction of the radiation 
flow in agreement with the results of simulations performed by \citet{lora09}. Taking into account the large distance (about 10 pc)
between the massive star and the YSO, this can be an important observational evidence to motivate the study
of this phenomena in such radiation source/YSO configuration.

These new results strengthen the suggestion of a star formation gradient presented in Paper\,I generated
by the action of the \hii~region G46, in which the formation processes have stopped at the pillars tips.

\section*{Acknowledgments}

We thank the anonymous referee for her/his very helpful comments and suggestions.
The ASTE project is led by Nobeyama Radio Observatory (NRO), a branch
of National Astronomical Observatory of Japan (NAOJ), in collaboration
with University of Chile, and Japanese institutes including University of
Tokyo, Nagoya University, Osaka Prefecture University, Ibaraki University,
Hokkaido University, and the Joetsu University of Education.
The INT is operated on the island of La Palma by the Isaac Newton Group in the Spanish Observatorio 
del Roque de los Muchachos of the Instituto de Astrof\'\i sica de Canarias. 
S.P., M.O., and A.P. are members of the {\sl Carrera del 
investigador cient\'\i fico} of CONICET, Argentina. M.C.P. is a doctoral fellow of CONICET, Argentina.
This work was partially supported by grants awarded by CONICET, ANPCYT and UBA (UBACyT) from Argentina.
A.P. aknowledges the support from the Varsavsky Foundation.
M.R. wishes to acknowledge support from FONDECYT(CHILE) grant N$^{\rm o}$1140839.
S.P. and A.P. are grateful to Dr. Takeshi Okuda for
the support received during the ASTE observations.

\label{lastpage}

\end{document}